\documentclass[pra,aps,reprint,letter,superscriptaddress,longbibliography,floatfix]{revtex4-2}
\usepackage{caption,subcaption}
\usepackage{times}
\usepackage{graphicx}
\usepackage{epsfig}
\usepackage{amsfonts}
\usepackage{amsmath}
\usepackage{amssymb}
\usepackage{color}
\usepackage{multirow}
\usepackage[colorlinks=true,linkcolor=blue,citecolor=magenta,urlcolor=blue]{hyperref}

\begin{document}
	

\title{Scalable estimation of pure multi-qubit states}


\author{L.~Pereira}
\email{luciano.ivan@iff.csic.es}
\affiliation{Instituto de F\'{\i}sica Fundamental IFF-CSIC, Calle Serrano 113b, Madrid 28006, Spain}

\author{L.~Zambrano}
\affiliation{Instituto Milenio de Investigaci\'on en \'Optica y Departamento de F\'{\i}sica, Universidad de Concepci\'on, casilla 160-C, Concepci\'on, Chile}

\author{A.~Delgado}
\affiliation{Instituto Milenio de Investigaci\'on en \'Optica y Departamento de F\'{\i}sica, Universidad de Concepci\'on, casilla 160-C, Concepci\'on, Chile}

	
\begin{abstract}
We introduce an inductive $n$-qubit pure-state estimation method. This is based on projective measurements on states of $2n$+1 separable bases or $2$ entangled  bases plus the computational basis. Thus, the total number of measurement bases scales as $O(n)$ and $O(1)$, respectively. Thereby, the proposed method exhibits a very favorable scaling in the number of qubits when compared to other estimation methods. Monte Carlo numerical experiments show that the method can achieve a high estimation fidelity. For instance, an average fidelity of $0.88$ on the Hilbert space of $10$ qubits is achieved with 21 separable bases. The use of separable bases makes our estimation method particularly well suited for applications in noisy intermediate-scale quantum computers, where entangling gates are much less accurate than local gates. We experimentally demonstrate the proposed method in one of IBM's quantum processors by estimating a 4-qubit Greenberger-Horne-Zeilinger states with a fidelity close to $0.875$ via separable bases. Other 10-qubit separable and entangled states achieve an estimation fidelity in the order of $0.85$ and $0.7$, respectively. 
\end{abstract}

\date{\today}

\pacs{Valid PACS appear here}
	
	
\maketitle

\section{Introduction}

In recent years, great advances have been achieved in the development of technologies based on quantum systems. These quantum technologies aim to provide significant performance improvements compared to their classical counterparts. 
    
A contemporary example is quantum computing, where large numbers of two-dimensional quantum systems (qubits) are controlled via canonical sets of operations to solve certain computational tasks more efficiently than classical computers \cite{Wright,Arute}. Several quantum systems, such as trapped ions \cite{Ions1,Ions2,Ions3,Ions4}, superconducting circuits \cite{Circuits1,Circuits2,Circuits3,Circuits4}, quantum dots \cite{dots1} and photonic platforms \cite{Multiports1,Multiports2,Multiports3,XANADU}, among others, have been proposed and tested to encode qubits and implement quantum gates, in order to build quantum simulators and quantum computers. Today, private companies provide cloud-based access to their quantum hardware and software. 
        
Currently, the above experimental platforms, among others, have evolved into devices that have been described as noisy intermediate-scale quantum (NISQ) technologies \cite{noise4}. These are devices that operate with imperfect gates on a number of qubits between 50 and up to a few hundred qubits. These devices are beyond the simulation capabilities of current classical computing devices but are still far from exhibiting a clear enough improvement in useful computing power. In particular, noisy quantum gates and decoherence limit the depth of the circuits and prevent the realization of complex algorithms \cite{noise1,noise2,noise3}.
    
Because of the properties of NISQ computers, the design of methods for the characterization, certification, and benchmarking of these quantum devices has become increasingly relevant and difficult \cite{review_characterization}. For instance, the characterization of states generated by this class of devices requires measurements on a large set of bases. These are physically implemented by applying sets of universal unitary transformations on the qubits followed by measurements on the computational basis. Typically, an effort to reduce the number of bases lead to a new set of bases that require the application of entangling unitary transformations, such as, for instance, the controlled NOT (CNOT) gate , whose implementation is associated to a large error. Consequently, the estimation of states generated by NISQ devices could be affected by a large build up of inaccuracy in the measurement procedure. 
    
Several methods have been proposed to assess the performance of NISQ devices. Randomized benchmarking \cite{RB1,RB2,RB3} and direct fidelity estimation \cite{Fidelity1,Fidelity2} are popular tools for this purpose. However, they do not give complete information about the device in question. In contrast, quantum state estimation \cite{review_QT} gives plenty of information about the device since it fully determines the state of the system from a set of suitable chosen measurements.

Today, several protocols to perform quantum state estimation of $d$-dimensional quantum systems are available. These attempt to estimate the $d^2-1$ real parameters that characterize density operators. Standard quantum tomography is based upon the measurement of the $d^2-1$ generalized Gell-Mann matrices \cite{SQT1,SQT2,SQT3}. Quantum state estimation based on projections onto the states of $d+1$ mutually unbiased bases has been suggested \cite{MUB1,MUB2,MUB3,MUB4,MUB5,MUB6,MUB7} to reduce the total number of measurement outcomes. This number can be reduced to a minimum using a symmetric, informationally complete, positive operator-valued measure \cite{SIC1,SIC2,SIC3,SIC4,SIC5,SIC6,SIC7}, which has exactly $d^2$ measurement outcomes.
	
Despite the progress made in the reduction of the total number of measurement outcomes used by quantum state estimation methods, the exponential increase of this number prevails in the case of multipartite systems. This is known as the \emph{curse of dimensionality}. For example, standard quantum tomography for a $n$-qubit system requires the measurement of $3^n$ local Pauli settings. A solution to overcome this problem is the use of {\it a priori} information about the state to be determined. Compressed sensing \cite{CS,Ahn} allows a large reduction in the total number of measurement as long as the rank of the state is known. Another alternative is to consider states with special properties, such as, for instance, matrix product states \cite{MPS} or permutationally invariant states \cite{Toth}, which also leads to a significant reduction in the total number of measurements. The important case of pure states has also been studied. In this case a set of five bases allows one to estimate with high accuracy all pure quantum states \cite{5B,CI5BB,5BFIJAS}. Recently, a 3-bases-based tomographic scheme has been introduced to estimate a single qudit \cite{3B}. The generalization of this method to multiple qubits leads to entangled bases. Adaptivity has also been introduced as a means of increasing the estimation accuracy of tomographic methods \cite {Mahler, ADAPTIVE, STRUCHALIN, Utreras, Fernandes} and even reaching fundamental accuracy limits \cite {Guo, Zambrano}.
	
Here, we present a method to estimate unknown pure quantum states in $n$-qubit Hilbert spaces, as used by NISQ computers, by means of $mn+1$ local bases (with $m$ an integer number), which correspond to the tensor product of $n$ single-qubit bases, or by the computational basis plus $m$ entangled bases, that is, bases that cannot be decomposed as the tensor product of single-qubit bases. Thereby, the number of bases scales linearly as $O(mn)$ and $O(m)$, respectively, with the number $n$ of qubits. This is a great advantage over standard methods, which are of exponential order. The $mn+1$ local bases also provide an advantage in the estimation of states generated via NISQ devices, such as current prototypes of quantum processors, because in this case projective measurements can be carried out without the use of typically noisy entangling gates. The computational basis together with $m=2$ entangled bases also lead to a clear reduction with respect to the 5-bases tomographic method \cite{5B,CI5BB}. The method here proposed also improves over the 3-bases method, since this requires to calculate the likelihood function of $2^{2^n-1}$ states to obtain an estimate. For a large number of qubits, this stage exponentially increases the computational cost of the method. Furthermore, in order to obtain distinguishable values of the likelihood, a large ensemble size is required. Our estimation method does not require such a procedure. 
	
The present method estimates pure multi-qubit quantum states using an inductive process. For an arbitrary $n$-qubit state $| \Psi \rangle$ we define $n$ sets of $2^{n-j}$ reduced states of dimension $2^{j}$, with $j=1, 2,..., n$ and with the reduced state of dimension $2^n$ equal to $| \Psi \rangle$ except a global phase. We first show how to estimate the reduced states of dimension $2$ and then show that the knowledge about reduced states of dimension $2^{j-1}$ together with some measurement outcomes allows one to estimate the reduced states of dimension $2^{j}$. This points to an inductive estimation method: we first estimate the reduced states of dimension $2$, and in $n$ iterations we arrive to the reduced state of dimension $2^n$, completing the estimation procedure. We show that the set of reduced states can be estimated by projecting the state onto $mn+1$ local bases or $m$ entangled bases plus the computational basis. Throughout the proposed estimation method, it is necessary to solve several systems of linear equations that might have vanishing determinants if the number of measurements is insufficient, leading to the failure of the method. However, the presence of noise, such as ,finite statistics, helps to mitigate this problem. We also show that a small increase in the number of local or entangled bases solves this problem and improves the overall fidelity of the estimation.
	
We study the present method through Monte Carlo numerical simulations. We randomly generate, according to a Haar-uniform distribution, a set of unknown pure states and calculate the average estimation fidelity as a function of the number of qubits. Projective measurements on each basis are simulated considering a fixed number of repetitions. We show that $2n+1$ separable bases achieve an average estimation fidelity of 0.88 for $n=10$. We improve this figure by considering larger number of bases, that is, $3n+1$ and $4n+1$, which lead to an average estimation fidelity of 0.915 and 0.93 for $n=10$, respectively. If the average estimation fidelity is calculated with respect to the set of separable states, then for the above number of bases we obtain the values $0.95, 0.955$ and $0.96$, correspondingly. The use of entangled bases leads to a less favorable picture, where $2,3$ and $4$ entangled bases plus the computational basis lead to an average estimation fidelity of 0.2, 0.6 and 0.8 for $n=6$, respectively. These figures increase when estimating separable states, where we achieve an average estimation fidelity of 0.75, 0.95 and 0.95 for $2,3$ and $4$ entangled bases plus the computational base, respectively. The large gap between separable and entangled bases can be explained by recalling that measurements are simulated with a fixed number of repetitions or, equivalently, with a fixed ensemble size. Thereby, the total ensemble size is much smaller in the case of the entangled bases, which decreases the estimation fidelity.
    
We test the proposed method by means of experiments carried out using the quantum processor \emph{ibmq\_manhattan} provided by IBM. We consider the estimation of fixed quantum states from 2 to 10 qubits using the local bases, and states up to 4 qubits using the entangled bases. We first estimate two completely separable states. One of these states is such that the matrices to be inverted are ill-conditioned. Nevertheless, the estimation of these states via local bases lead to very similar fidelities above 0.83 for $n=10$. The estimation via entangled bases leads to fidelities in the interval $[0.1,0.3]$ for $n=4$. We also consider the estimation of an $n/2$-fold tensor product of a Bell state, where local bases for $n=10$ lead to a fidelity above $0.67$ and entangled bases lead to a fidelity below 0.5 for $n=10$. Finally, we test the estimation of a Greenberger-Horne-Zeilinger state for 2, 3 and 4 qubits. The proposed method leads to en estimation fidelity of approximately 0.87 for $n=4$ via local bases while in the case of entangled bases the fidelity is approximately 0.81, 0.53 and 0.47 for 2, 3 and 4 entangled bases plus the computational base, respectively. The large decrease in the estimation of entangled states by means of entangled bases can be explained by the use of low-accuracy entangling gates in the preparation of the states as well in the projection to the states of the entangled bases.
    
Our simulations and experimental results indicate that the method proposed here allows the efficient and scalable estimation of $n$-qubit pure states by means of $mn+1$ separable bases in NISQ computers, while the $m$ entangled bases together with the computational base can offer a large advantage in quantum computers based on high accuracy entangling gates.
    
This article is organized as follows: in Section II we present the method and its properties. In Section III we show the results of several Monte Carlo simulations aimed at studying the overall behavior of the fidelity as a function of the number of qubits. In Section IV we present the results of implementing the proposed estimation method in IBM's quantum processors. In Section V we summarize and conclude.

\section{Method}
    
Let us consider a $n$-qubit system described by the pure state
\begin{align}
|\Psi\rangle = \sum_{\alpha=0}^{2^n-1} c_\alpha e^{i\phi_\alpha} |\alpha\rangle_n,
\end{align}
where $c_\alpha\geq0$ and $|\alpha\rangle_n =|\alpha_{n-1}\rangle\otimes\cdots\otimes|\alpha_0\rangle$ is the $n$-qubit computational basis, with $\alpha = \sum_{k=0}^{n-1}2^k\alpha_k $ the integer associated with the $n$-bit binary string $\alpha_{n-1}\cdots \alpha_0$. 
    
Our main aim is to estimate the values of the amplitudes $\{c_\alpha\}$ and the phases $\{\phi_\alpha\}$ with a total number of measurements that does not scale exponentially with the number $n$ of qubits. For this purpose, we employ an iterative algorithm based on estimating reduced states, which are non-normalized $j$-qubit vectors defined by 
\begin{align}
|\Psi_{\beta}^j\rangle =\sum_{\alpha=0}^{2^j-1} {}_{n}\langle 2^{n-j}\beta + \alpha|\Psi\rangle |\alpha\rangle_j,
\end{align}
with $1\leq j \leq n$ and $ 0 \leq \beta \leq 2^{n-j} -1$. Thus, for a given state $| \Psi\rangle$ we have $n$ sets of $2^{n-j}$ reduced states, one set for each $j$ and one reduced state for each $\beta$. 

Every reduced state has some partial information about the full state. For instance, if $j=1$ we have $2^{n-1}$ reduced states that are given by
\begin{align}
|\Psi_{\beta}^1\rangle = c_{2\beta}e^{i\phi_{2\beta}}|0\rangle + c_{2\beta+1}e^{i\phi_{2\beta+1}}|1\rangle. \label{Psi_j1}
\end{align}
We can see that the amplitudes and the phases are the same entering in the state $| \Psi \rangle$.  Thereby, we can reduce the problem of estimating the unknown state $|\Psi\rangle$ to the problem of estimating the set of reduced states. The main obstacle of this approach is that measurements in quantum mechanics do not contain information about global phases, and hence, if we try to find every reduced state in an independent way using the results of some measurements, we can only obtain the reduced states $|\Psi_{\beta}^j\rangle$ except a global phase, which in this case, would be a relative phase in $|\Psi \rangle$. Thus, the only reduced state that fully characterizes the system is $|\Psi_{0}^n\rangle$, which is the same as $|\Psi \rangle$. Nevertheless, the remaining reduced states will also be useful in the task of reconstructing $|\Psi \rangle$, because we can find the reduced states $| \Psi_\beta^{j} \rangle$ (except a global phase) using the reduced states $| \Psi_\beta^{j-1} \rangle $, as we will show.

For a large enough set of identical copies of the state $|\Psi\rangle$, measurements in the computational base $\{| \alpha \rangle \}$ gives us a histogram of observations such that we have approximations $p_\alpha$ to $|\langle\alpha|\Psi\rangle|^2$. Then the amplitudes can be estimated as $c_{\alpha} = \sqrt{p_{\alpha}}$. On the other hand, the phases can be determined from projective measurements on the following states
\begin{align}
|P_{a\beta}^j\rangle = |\beta\rangle_{n-j}\otimes|+_a\rangle\otimes|-_a\rangle^{\otimes j-1}, \label{ProjTomo}
\end{align}
where $|+_a\rangle = u_{a}|0\rangle + v_{a}e^{i\varphi_a}|1\rangle$ and $|-_a\rangle = v_{a}|0\rangle - u_{a}e^{i\varphi_a}|1\rangle$ are orthonormal single-qubit states, with $a = 1,\dots, m$. Here, $m$ is the number of bases $\{ |\pm_a\rangle\}$ considered, which must be large enough to carry out the algorithm. In addition, these bases have to be all different from the computational one.

To start the algorithm, we set $j=1$ and the reduced states are given by Eq. \eqref{Psi_j1}. The probabilities of projecting $|\Psi\rangle$ onto $|P^1_{a\beta}\rangle$ are given by
\begin{align}
P^1_{a\beta} 
=& u_{a}^2c_{2\beta}^2 + v_{a}^2c_{2\beta+1}^2 \nonumber \\
 & + 2u_{a}v_{a}c_{2\beta}c_{2\beta+1} [ \cos(\varphi_a)\cos(\phi_{2\beta+1}-\phi_{2\beta}) \nonumber \\
 & +\sin(\varphi_a)\sin(\phi_{2\beta+1}-\phi_{2\beta}) ].
\label{P1abeta}
\end{align}
These generate a set of equations, one equation for each value of $\beta$, that are linear combinations of the cosine and sine of the relative phase $\delta\phi_\beta = \phi_{2\beta+1}-\phi_{2\beta}$, with coefficients depending on $|+_a\rangle$. Thereby, we can form a linear system of equations $L_\beta \vec{a}_\beta = \vec{b}_\beta$ for trigonometric functions of the relative phase, which is explicitly given by
\begin{align}
\begin{bmatrix}
\cos\varphi_1 & \sin\varphi_1 \\
\vdots & \vdots \\
\cos\varphi_m & \sin\varphi_m 
\end{bmatrix}
\begin{bmatrix}
\cos\delta\phi_\beta\\
\sin\delta\phi_\beta  
\end{bmatrix} 
= 
\begin{bmatrix}
\tilde{P}^{1}_{1\beta}\\
\vdots \\
\tilde{P}^{1}_{m\beta}
\end{bmatrix},\label{EqSyst1}
\end{align} 
where we have defined
\begin{align}
\tilde{P}_{a\beta}^{1} = \frac{1}{2u_{a}v_{a}c_{2\beta}c_{2\beta+1} }(P_{a\beta}^{1} - u_{a}^2c_{2\beta}^2 - v_{a}^2c_{2\beta+1}^2 ). 
\end{align}
Since we know the projectors $|P^1_{a\beta}\rangle$ and the coefficients $c_j$, this system can be solved for $m\geq 2$ inverting $L_\beta$ by the Moore-Penrose pseudo-inverse $A^+=(A^TA)^{-1}A^T$. We can guarantee that the pseudo-inversion is possible by suitably choosing the coefficients $u_a$, $v_a$ and the phases $\varphi_a$ entering in the states $|P_{a\beta}^1\rangle$. For example, taking the particular case of $m=2$ the system has a solution 
\begin{align}
\begin{bmatrix}
\cos \delta\phi_\beta\\
\sin \delta\phi_\beta  
\end{bmatrix}
=\frac{1}{\det(L_\beta)}
\begin{bmatrix}
\sin \varphi_2  & -\sin \varphi_1 \\
-\cos \varphi_2 & \cos \varphi_1
\end{bmatrix}
\begin{bmatrix}
\tilde{P}^{1}_{1\beta}\\
\tilde{P}^{1}_{2\beta}
\end{bmatrix}
\end{align}
as long as the determinant of $L_\beta$ does not vanish, that is, $\det(L_\beta)=\cos(\varphi_1)\sin(\varphi_2)-\sin(\varphi_1)\cos(\varphi_2) \neq 0$. Thereby, taking $\varphi_1=0$ and $\varphi_2=\pi/2$, or equivalently $|\pm_1\rangle=u_1|0\rangle\pm v_1|1\rangle$, $|\pm_2\rangle=u_2|0\rangle\pm i v_2|1\rangle$, the system of equations can be always solved since $\det(L_\beta)=1$, and the exponential of the relative phase is given by $e^{i\delta\phi_\beta} = \tilde{P}^1_{1\beta} + i\tilde{P}^1_{2\beta}$. Therefore, the reduced states $|\Psi_{\beta}^1\rangle$ can be determined up to a global phase,
\begin{align}
|\tilde\Psi_{\beta}^1\rangle = c_{2\beta}|0\rangle + c_{2\beta+1}e^{i\delta\phi_\beta}|1\rangle, \label{Eq:ReducedStatej1}
\end{align}
provided we measure the computational basis and at least two projectors $|P_{a\beta}^1 \rangle$ for every $\beta$ in $0, 1,..., 2^{n-1}-1$. Nevertheless, when some of the computational basis probabilities $p_\alpha$ are null, we have to consider a particular rule to obtain the reduced state. If any or both of the coefficients $c_{2\beta}=\sqrt{p_{2\beta+1}}$ or $c_{2\beta+1}=\sqrt{p_{2\beta+1}}$ are null, the reduced state is determined only by the computational basis, setting $|\tilde\Psi_{\beta}^1\rangle = c_{2\beta} | 0 \rangle$, $|\tilde\Psi_{\beta}^1\rangle = c_{2\beta + 1} | 1 \rangle$ or $|\tilde\Psi_{\beta}^1\rangle = \vec{0}$ in Eq. \eqref{Eq:ReducedStatej1}. This means that there may be null reduced states, which will have an impact on the algorithm in posterior iterations.  

In the following iterations, that is, the case $j>1$, the reduced states can be expressed as linear combinations of the previous ones,
\begin{align}
|\tilde\Psi_{\beta}^j\rangle = |0\rangle\otimes|\tilde\Psi_{2\beta}^{j-1} \rangle + e^{i\delta\phi_\beta^j}|1\rangle\otimes|\tilde\Psi_{2\beta+1}^{j-1} \rangle,
\end{align}
with $|\tilde\Psi_{\beta}^j\rangle = e^{-i\phi_{2^j\beta}}|\Psi_{\beta}^j\rangle$ the corresponding reduced states up to global phase and $\delta\phi_\beta^j =\phi_{2^{j}(\beta+1/2)}-\phi_{2^{j}\beta} $ relative phases. Thus, assuming that we know the reduced states of the previous iteration (except for a global phase), we can determine the reduced state of the current iteration (except for a global phase) simply by determining the relative phase. Analogously to the first iteration, if any or both of the previous reduced states $|\tilde\Psi_{2\beta}^{j-1} \rangle$ or $|\tilde\Psi_{2\beta+1}^{j-1} \rangle$ are null, there is no relative phase to determine and the next reduced state is simply obtained with the choice $\delta\phi_{\beta}=0$. Otherwise, we have to determine the relative phase from the projections $|P_\beta^j\rangle$. Considering $j>1$, the probability of projecting the state $|\Psi\rangle$ onto the state $|P^j_{a\beta}\rangle$ is given by
\begin{align}
P^j_{a\beta} 
=& u_{a}^2 |\langle W_a^j | \tilde\Psi_{2\beta}^{j-1} \rangle|^2 +v_{a}^2 |\langle W_a^j | \tilde\Psi_{2\beta+1}^{j-1} \rangle|^2 \nonumber \\
& + 2u_{a}v_{a}\Re[ e^{i(\delta\phi_\beta^j-\varphi_a)} \langle \tilde\Psi_{2\beta}^{j-1} |W_a^j \rangle\langle W_a^j | \tilde\Psi_{2\beta+1}^{j-1} \rangle     ],
\end{align} 
where $|W_a^j\rangle = | -_a\rangle^{\otimes j-1}$. Defining the quantities 
\begin{align}
X_{a\beta}^j = e^{-i\varphi_a} \langle \tilde\Psi_{2\beta}^{j-1} |W_a^j \rangle\langle W_a^j | \tilde\Psi_{2\beta+1}^{j-1} \rangle 
\end{align}
and 
\begin{align}
\tilde{P}_{a\beta}^j = \frac{1}{2u_{a}v_{a} }\big( P_{a\beta}^j - u_{a}^2 |\langle W_a^j | \tilde\Psi_{2\beta}^{j-1} \rangle|^2 - v_{a}^2 |\langle W_a^j | \tilde\Psi_{2\beta+1}^{j-1} \rangle|^2  \big),
\end{align}
we obtain the following system of equations for the relative phases
\begin{align}
\begin{bmatrix}
\Re X_{1\beta}^j & -\Im X _{1\beta}^j\\
\vdots & \vdots \\
\Re X_{m\beta}^j & -\Im X _{m\beta}^j
\end{bmatrix}
\begin{bmatrix}
\cos\delta\phi_\beta^j\\
\sin\delta\phi_\beta^j
\end{bmatrix}
=
\begin{bmatrix}
\tilde{P}_{1\beta}^j\\
\vdots \\
\tilde{P}_{m\beta}^j
\end{bmatrix}, \label{EqSyst}
\end{align}
where $\Re$ and $\Im$ stand for real and imaginary part, respectively. Again, the phases can be obtained by solving a linear system of equations $L_{\beta}^j\vec{a}_{\beta}^j=\vec{b}_{\beta}^j$ given by Eq. \eqref{EqSyst} through the Moore-Penrose pseudo-inverse. If for a fixed $j$, we are able to pseudo-invert the matrices $L_{\beta}^j$, we can find all the $|\tilde\Psi_{\beta}^j\rangle$. Then, we continue inductively with $j+1$ until reaching $j=n$. In that case, the protocol ends because the reduced state $|\tilde\Psi_{0}^n\rangle$ is equal to the state $|\Psi\rangle$ up to a global phase. 

One might think that, analogously to the case $j = 1$, the matrix $L_{\beta}^j$ can be inverted for $m\geq 2$. Thereby, to completely estimate the unknown state $|\Psi\rangle$ we would need a minimum of $2(2^n-1)$ different projections onto states $|P_{a\beta}^j\rangle$ in \eqref{ProjTomo}, because there are $2^n-1$ different reduced states. However, now the matrix $L_{\beta}^j$ to be pseudo-inverted does not only depends on the projectors $|P^j_{a\beta}\rangle$, but it also depends on the reduced states of the previous iteration, so that we cannot guarantee that the pseudo-inversion is feasible. For example, taking $m=2$, the solution of the system of equations \eqref{EqSyst} is
\begin{align}
\begin{bmatrix}
\cos\delta\phi_\beta^j\\
\sin\delta\phi_\beta^j
\end{bmatrix}
=
\frac{1}{\det(L_\beta^j)}\begin{bmatrix}
-\Im X _{2\beta}^j& \Im X_{1\beta}^j\\
-\Re X_{2\beta}^j & \Re X_{1\beta}^j 
\end{bmatrix}
\begin{bmatrix}
\tilde{P}_{1\beta}^j\\
\tilde{P}_{2\beta}^j
\end{bmatrix},
\label{EQ-SYS}
\end{align}
with $\det(L_\beta^j)=\Im(X_{1\beta}^j[X_{2\beta}^j]^*)$ the determinant of $L_\beta^j$. Clearly, this solution is only valid when $\det(L_\beta^j)\neq 0$. A non-invertible system of equations for a $|\tilde\Psi_{\beta}^j\rangle$ means that all the equations are linearly dependent and that it is sufficient to consider a single equation $\Re X_{\beta}^j\cos(\delta\phi_{\beta}^j)-\Im X_{\beta}^j\sin(\delta\phi_{\beta}^j) = \tilde{P}_{\beta}^j$, from which the trigonometric functions of the relative phase cannot be solved. However, using the identity $\cos^2(\delta\phi_{\beta}^j)+\sin^2(\delta\phi_{\beta}^j)=1$, we can obtain the trigonometric functions except for their corresponding quadrant, that is, we have two possible estimates or ambiguities for the reduced states. Thus, in the worst case, where the matrix $L_{\beta}^j$ cannot be pseudo-inverted for any reduced state, the algorithm leads to a finite set of $2^{n-1}$ possible estimates of $|\Psi\rangle$. This problem can be avoided by increasing the number of $m$ of bases for given values of $j$ and $\beta$ until the systems can be solved. An alternative solution is to consider an extra adaptive measurement to discriminate between the ambiguities. Furthermore, since the probabilities are experimentally estimated by means of a sample of finite size, with high probability the system of equations can be solved due the finite sample noise, although not necessarily with high accuracy since $L_{\beta}^j$ might have a bad condition number, $\textrm{cond}(A) = ||A|| \times ||A^+||$, with $||\cdot||$ a matrix norm.  

\begin{figure}
    \begin{subfigure}[b]{0.5\textwidth}
    \centering
    \includegraphics[scale=0.32]{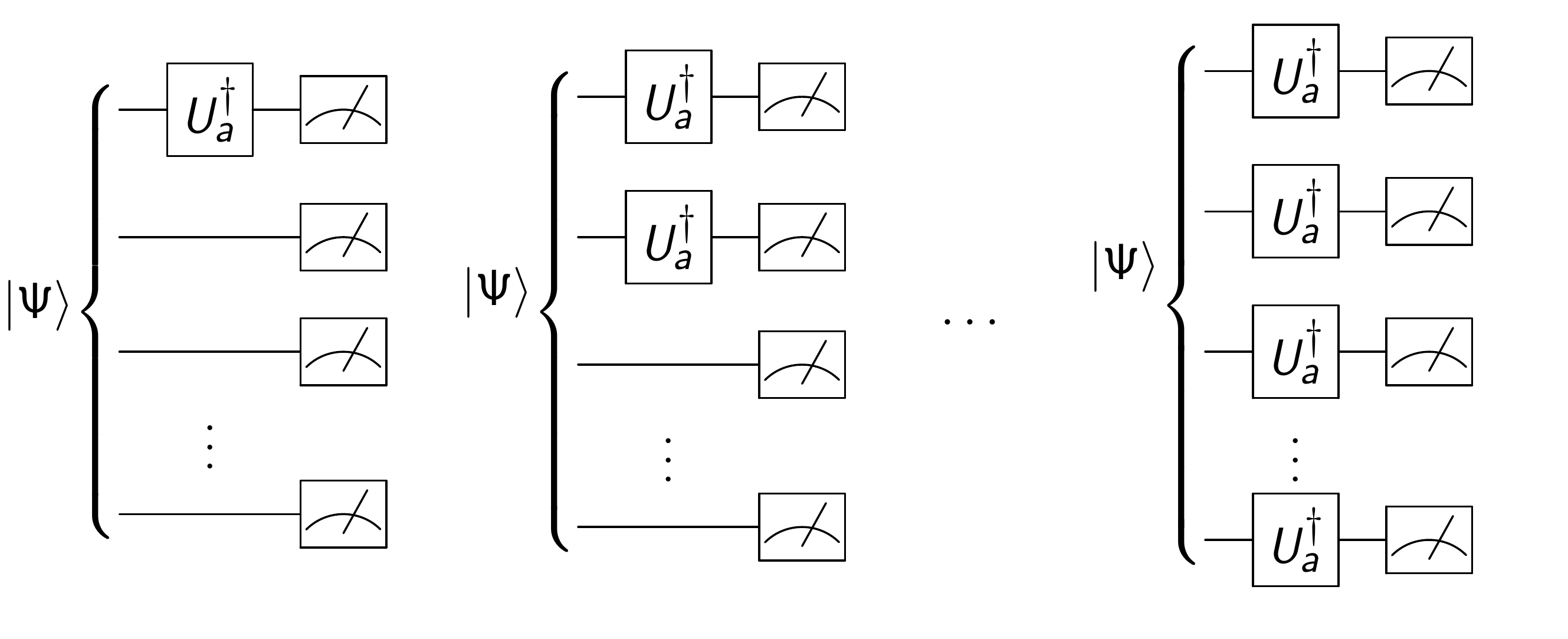}
    \caption{Local basis.}
    \label{circ_local}
    \end{subfigure}
    \begin{subfigure}[b]{0.5\textwidth}
    \centering
    \includegraphics[scale=0.3]{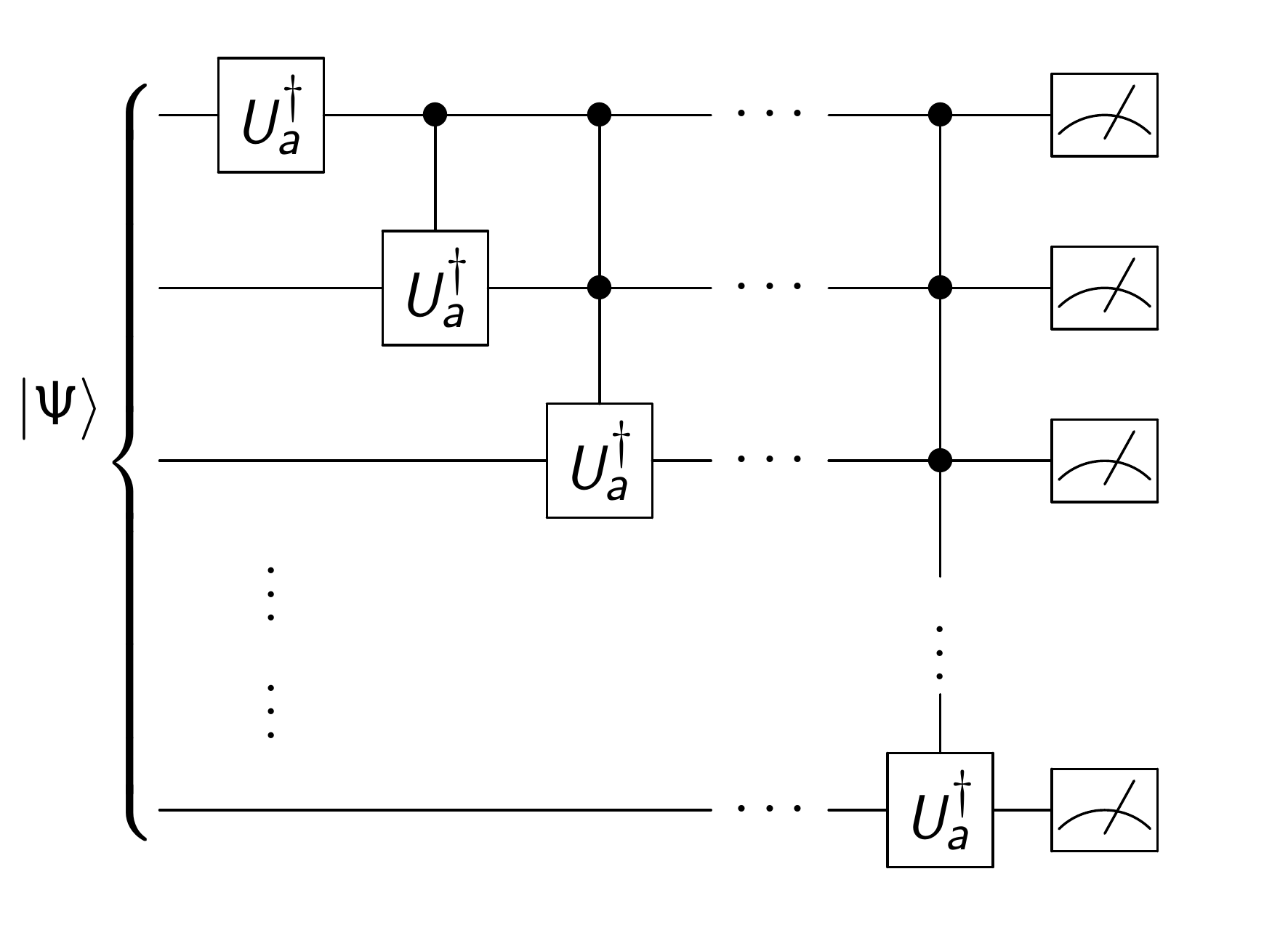}
    \caption{Entangled basis.}
    \label{circ_entangled}
    \end{subfigure}
    \caption{Quantum circuits to implement \ref{circ_local} the local measurements $\{\mathcal{L}_{ab}\}$ and \ref{circ_entangled} the entangled measurements $\{\mathcal{E}_a\}$. The single qubit gate $U_a$ fulfill $|+_a\rangle=U_a|0\rangle$ and $|-_a\rangle=U_a|1\rangle$.}
    \label{fig:circuits}
\end{figure}

In order to apply our method based on the inductive estimation of reduced states we need to find an efficient implementation of projections onto states $|P_{a\beta}^j\rangle$. For this purpose we consider the measurement of two different sets of bases. The first set of $n\times m$ bases are given by
\begin{equation}
\mathcal{L}_{ab} = \left\{|\beta\rangle_{n-b}\otimes |\pm_a\rangle \otimes \cdots \otimes |\pm_a \rangle  \right\}, \label{local_observables}
\end{equation}
with $ 0 \leq \beta \leq 2^{n-b} -1$. These bases can be implement via local gates, as shown in Fig.~\ref{fig:circuits}. These bases define a larger set of projections than required by the method. However, these extra projections can also be used in the algorithm, since $|\beta\rangle_{n-j}\otimes |\pm_a\rangle \otimes \cdots \otimes |\pm_a \rangle $ form a system of $2^{j}$ equations at the $j$th-iteration. The second set of $m$ bases are given by
\begin{equation}
\mathcal{E}_a = \left\{  |\beta\rangle_{n-j}\otimes |+_a\rangle \otimes |-_a\rangle ^{\otimes j-1}, |-_a\rangle^{\otimes n}  \right\}, \label{ent_observables}
\end{equation}
with $1\leq j \leq n$ and $ 0 \leq \beta \leq 2^{n-j} -1$. Despite projections onto members of the bases are local, these observables are entangled since many multi-controlled $U$ gates are needed to implement them, as shown in Fig.~\ref{fig:circuits}. In a quantum computer, these gates can be decomposed in terms of CNOT gates \cite{MultiControlGates}. 

As example, let us consider a 2-qubit system in the state
\begin{align}
    |\Psi\rangle = c_0e^{i\phi_0}|0\rangle_2 + c_1e^{i\phi_1}|1\rangle_2 +c_2e^{i\phi_2}|2\rangle_2 +c_3e^{i\phi_3}|3\rangle_2. 
\end{align}
Omitting the Kronecker product for compactness, the corresponding reduced states are 
\begin{align}
    |\tilde\Psi^1_0\rangle &= c_0|0\rangle + c_1e^{i(\phi_1-\phi_0)}|1\rangle,\\
    |\tilde\Psi^1_1\rangle &= c_2|0\rangle +c_3e^{i(i\phi_3-\phi_2)}|1\rangle,\\
    |\tilde\Psi^2_0\rangle &= |0\rangle\big(c_0|0\rangle + c_1e^{i(\phi_1-\phi_0)}|1\rangle \big) \nonumber \\
    &\quad + e^{i(\phi_2-\phi_0)} |1\rangle\big( c_2|0\rangle +c_3e^{i(i\phi_3-\phi_2)}|1\rangle \big), 
\end{align}
and, the bases to be measured are
\begin{align}
    \mathcal{L}_{a0} =& \{\: |0\rangle|+_a\rangle,\: |0\rangle|-_a\rangle,\: |1\rangle|+_a\rangle,\: |1\rangle|-_a\rangle \:  \},\\
    \mathcal{L}_{a1} =& \{\: |+_a\rangle|+_a\rangle,\: |+_a\rangle|-_a\rangle, \:|-_a\rangle|+_a\rangle, \:|-_a\rangle|-_a\rangle \:  \} ,
\end{align}
or
\begin{align}
    \mathcal{E}_a =\{\: |0\rangle|+_a\rangle,\: |1\rangle|+_a\rangle,\: |+_a\rangle|-_a\rangle,\: |-_a\rangle|-_a\rangle\: \}.
\end{align}
 
\begin{figure*}[t!]
  \begin{subfigure}[b]{0.245\textwidth}
    \includegraphics[width=\textwidth]{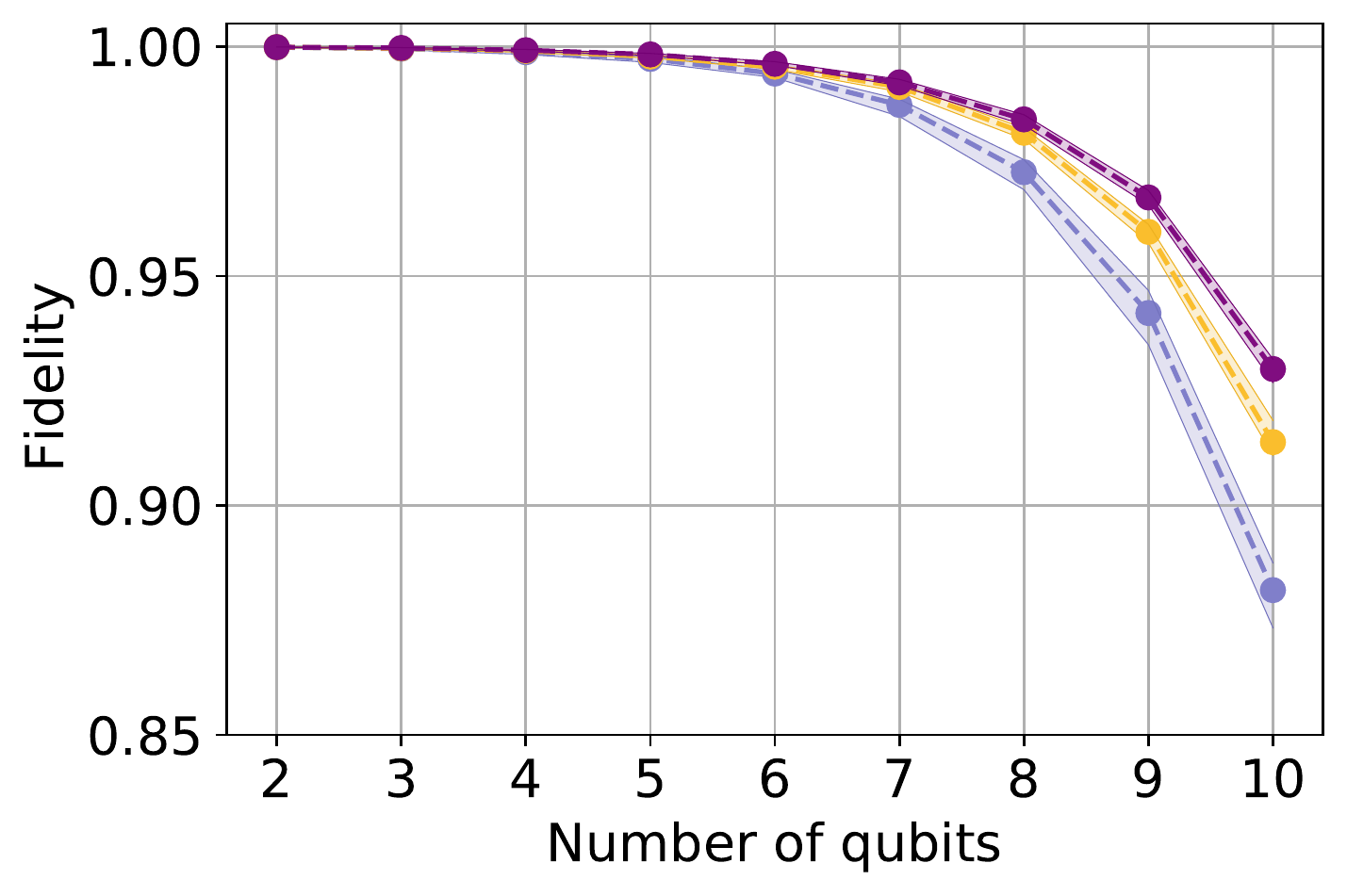}
    \caption{Randomly generated states, local bases}
    \label{fig:sim_2_to_10_ent}
  \end{subfigure}
  \begin{subfigure}[b]{0.245\textwidth}
    \includegraphics[width=\textwidth]{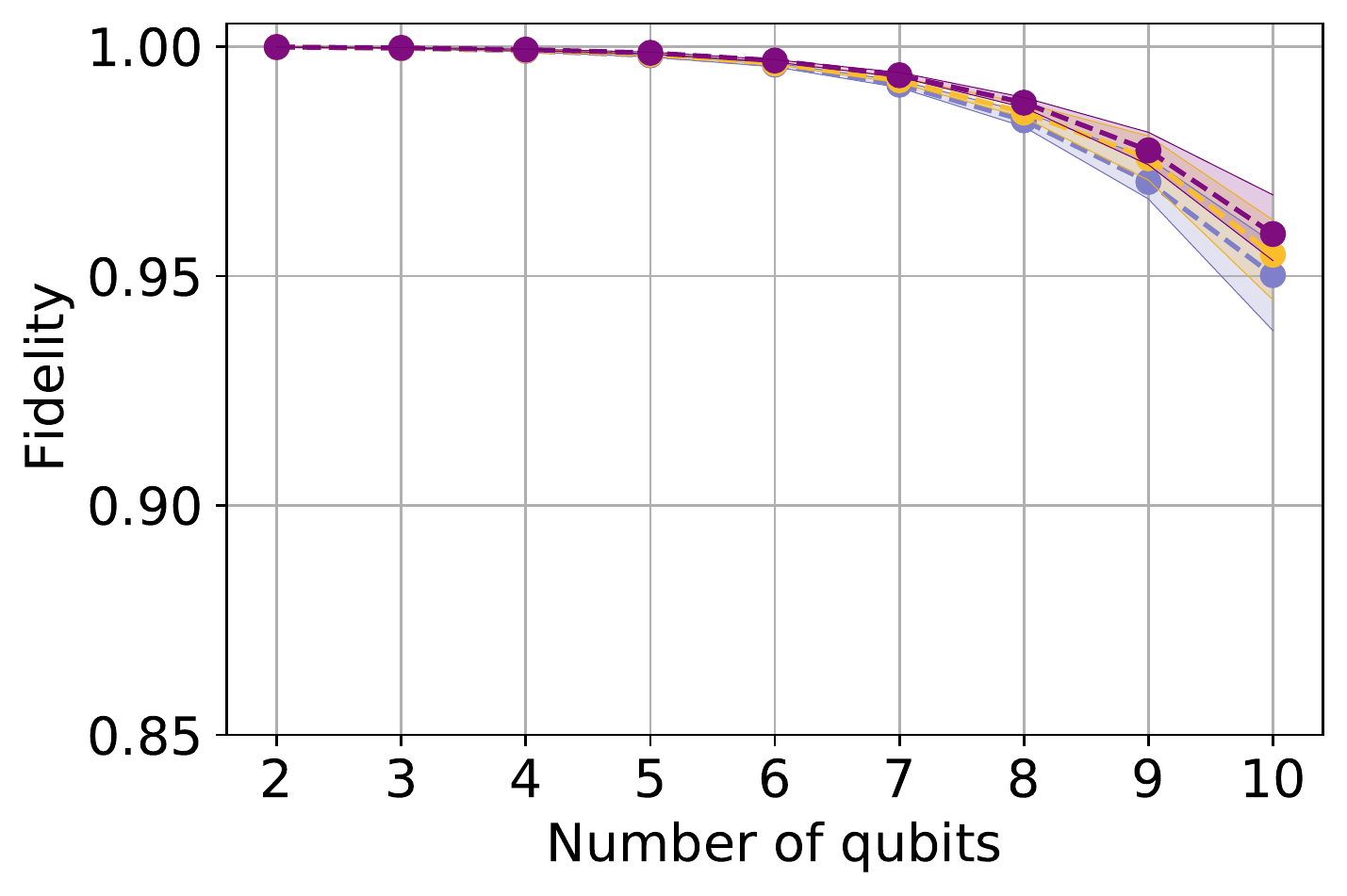}
    \caption{Randomly generated separable states, local bases}
    \label{fig:sim_2_to_10_sep}
  \end{subfigure}
    \begin{subfigure}[b]{0.245\textwidth}
    \includegraphics[width=\textwidth]{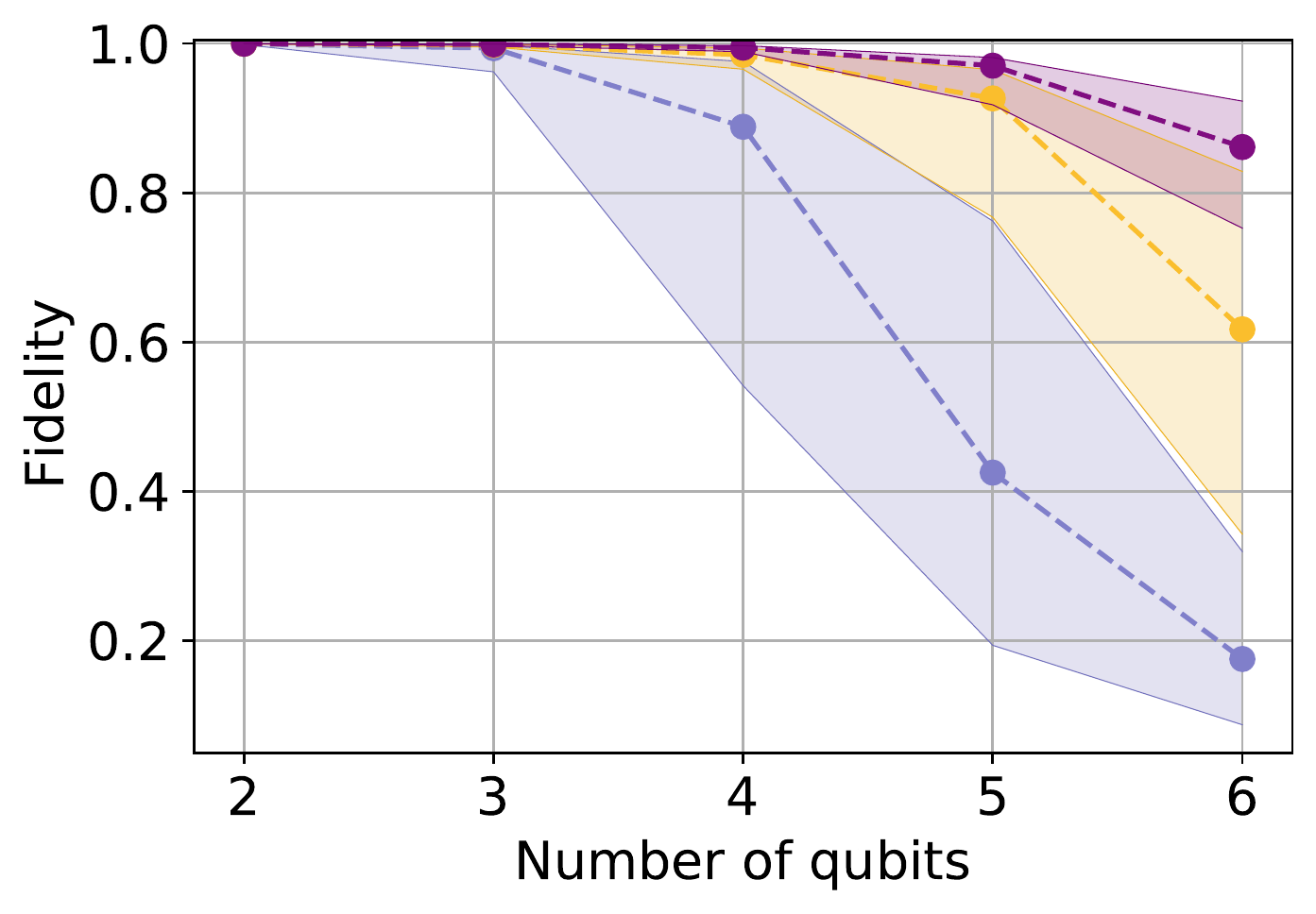}
    \caption{Randomly generated states, entangled bases}
    \label{fig:sim_2_to_6_ent_ent}
  \end{subfigure}
  \begin{subfigure}[b]{0.245\textwidth}
    \includegraphics[width=\textwidth]{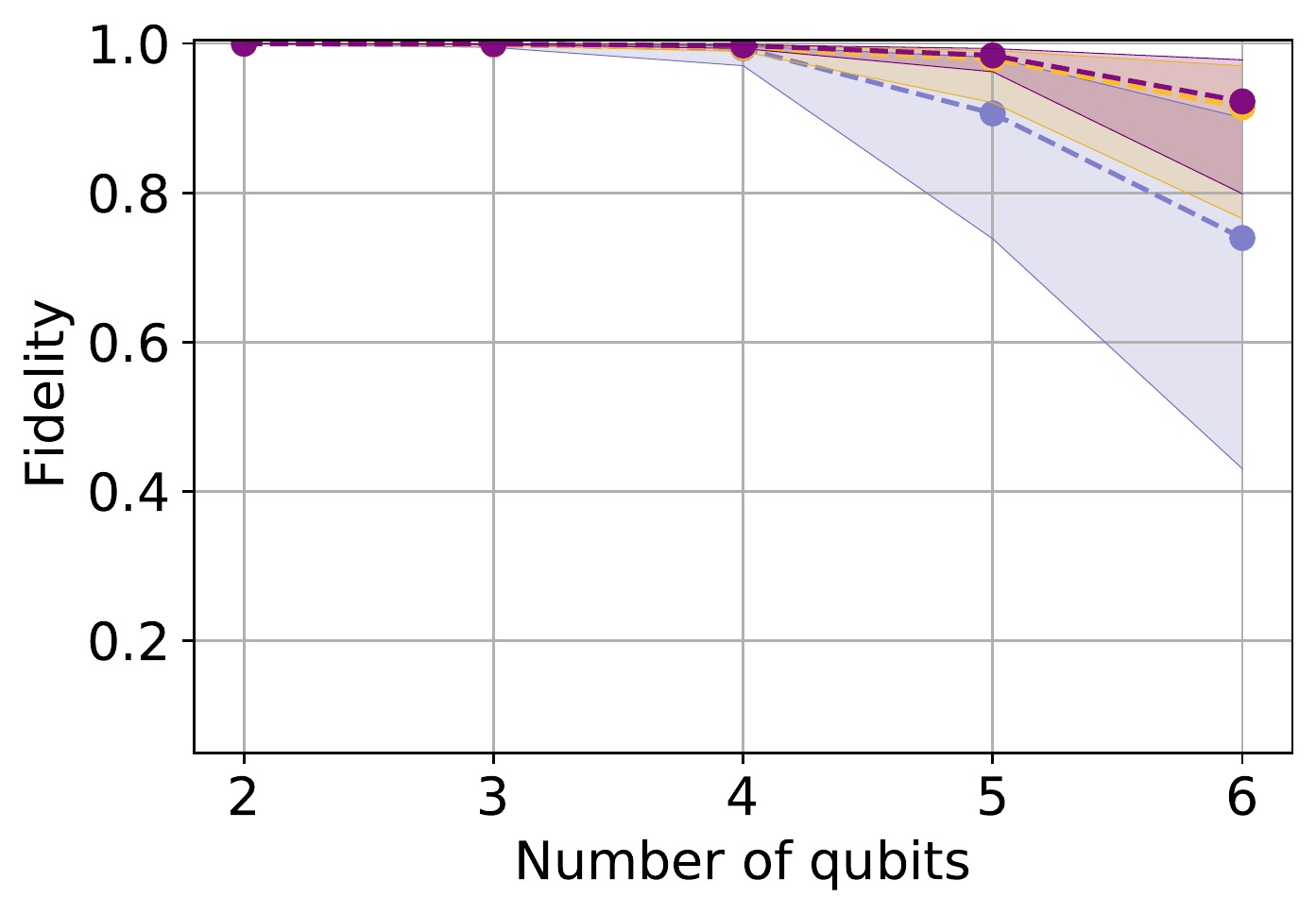}
    \caption{Randomly generated separable states, entangled bases}
    \label{fig:sim_2_to_6_sep_ent}
  \end{subfigure}
\caption{Median fidelity (solid dots) between arbitrary states and its estimates as a function of number of qubits. The colors label the number of bases: For (a) and (b) we have $2n+1$ (blue), $3n+1$ (yellow) and $4n+1$ (purple) local bases. For (c) and (d) we have $2$ (blue), $3$ (yellow) and $4$ (purple) entangled bases plus the computational basis. Shaded areas represent the corresponding interquartile range.}
\label{Figure2}
\end{figure*}

\section{Numerical Simulations}

We are interested in how similar an unknown state and its estimate are. To quantify this we use the fidelity defined by 
\begin{align}
    F = | \langle \Psi | \Psi_{est} \rangle |^2,
\end{align}
between an arbitrary state $| \Psi \rangle $ and its estimate $|\Psi_{est} \rangle$. A vanishing fidelity indicates that $|\Psi\rangle$ and $|\Psi_{est}\rangle$ can be perfectly distinguished. A unitary fidelity indicates that $|\Psi\rangle$ and $|\Psi_{est}\rangle$ are equal. Thus, a good pure-state estimation method will be characterized by high values of the fidelity.
We are interested in the behavior of the fidelity as a function of the dimension, or equivalently, the number of qubits, and the impact on the fidelity of shot noise or finite-statistics effects.

To study the performance of the proposed method we carried out numerical experiments. For each number of qubits $n=2,3,\dots,10$ we randomly generate a set $\{|\Psi^{(i)}\rangle\}$ of $100$ Haar-distributed pure states, which play the role of the unknown state to be estimated. Afterward, we simulate projective measurements on each $|\Psi^{(i)}\rangle$ using the entangled bases, Eq.~\eqref{ent_observables}, and the local bases, Eq.~\eqref{local_observables}. These measurements are simulated considering $10^{13}$ repetitions for each base. Thereafter, using the set of estimated probabilities obtained from the simulated measurements we applied our estimation method to each state $|\Psi^{(i)}\rangle$, which leads to a set $\{|\Psi_{est}^{(i)} \rangle\}$ of estimates. Finally, we calculate the value of the fidelity between each unknown state and its estimate. Thereby, we obtain a set of fidelity values for each number $n$ of qubits, the statistics of which we study in the following figures.

Figure $\ref{fig:sim_2_to_10_ent}$ shows the median of the fidelity as a function of the number $n$ of qubits for three different numbers $nm + 1$ of local bases with  $m=2, 3$, and $4$, from bottom to top. For each number of bases, the infidelity decreases as the number of qubits increases. This is a general feature of quantum state estimation methods. Since for each basis the number of repetitions is fixed, the probabilities entering in the systems of linear equations, such as Eq.~(\ref{EQ-SYS}), are estimated with a decreasing accuracy as $n$ increases, which leads to lower fidelity values. Figure $\ref{fig:sim_2_to_10_ent}$ also shows that the median value of the fidelity can be increased, while keeping the number of repetitions fixed, using a larger number of local bases. The proposed method requires the pseudo-inverse for every matrix in Eq.~\eqref{EqSyst}. The quality of this inversion can be improved by reducing the condition numbers of the linear equation systems, which can be achieved by increasing the number of bases employed in the estimation. An interesting feature of Fig. ~$\ref{fig:sim_2_to_10_ent}$ is that the interquartile range is very narrow for each basis, which indicates that the estimation method generates very similar fidelity values for all the simulated states.

Figure \ref{fig:sim_2_to_6_ent_ent} exhibits the median of the fidelity as a function of the number $n$ of qubits for 2, 3 and 4 entangled bases plus the canonical basis. In this case, the estimation method achieves a poor performance in comparison to the use of local bases. The use of 2 entangled bases lead to two equations for every reduced state, which generates an ill-conditioned equation system. Thereby, a poor estimation of the phases of the probability amplitudes is obtained, which propagates through the iterations of the algorithm leading to a low fidelity. However, as Fig.~\ref{fig:sim_2_to_6_ent_ent} shows, the fidelity can be greatly improved by increasing the number of entangled bases for the estimation. Also, the simulation was carried out employing a fixed ensemble size of $2^{13}$ for each base. This was done to allow a comparison between our simulations and the results of experiments on the IBM's quantum processors. The total ensemble employed with the local and entangled bases scales as $mn\times2^{13}$ and $k\times2^{13}$, respectively. Thereby, the total ensemble used with the local bases is much larger than in the case of the entangled bases, which leads to a better estimation via local bases. An increase of the ensemble size used with the entangled bases could lead to an improvement of the fidelity. Nevertheless, the entangled bases are capable of delivering for $n=6$ a fidelity close to $0.9$. 

Figures \ref{fig:sim_2_to_10_sep} and \ref{fig:sim_2_to_6_sep_ent} also display the fidelity as a function of the number of qubits for local and entangled bases, respectively. However, the simulation has been carried out on the set of completely separable states. In both cases we obtain an increase of the performance of the proposed estimation method, achieving fidelities of around $0.95$ for $10$ qubits when employing local bases. This result indicates that the performance of the proposed estimation method on the set of entangled states should be closer to the depicted in Figs. $\ref{fig:sim_2_to_10_ent}$ and \ref{fig:sim_2_to_6_ent_ent}.

\begin{figure*}[t!]
\begin{subfigure}{.245\textwidth}
  \centering
  \includegraphics[width=\textwidth]{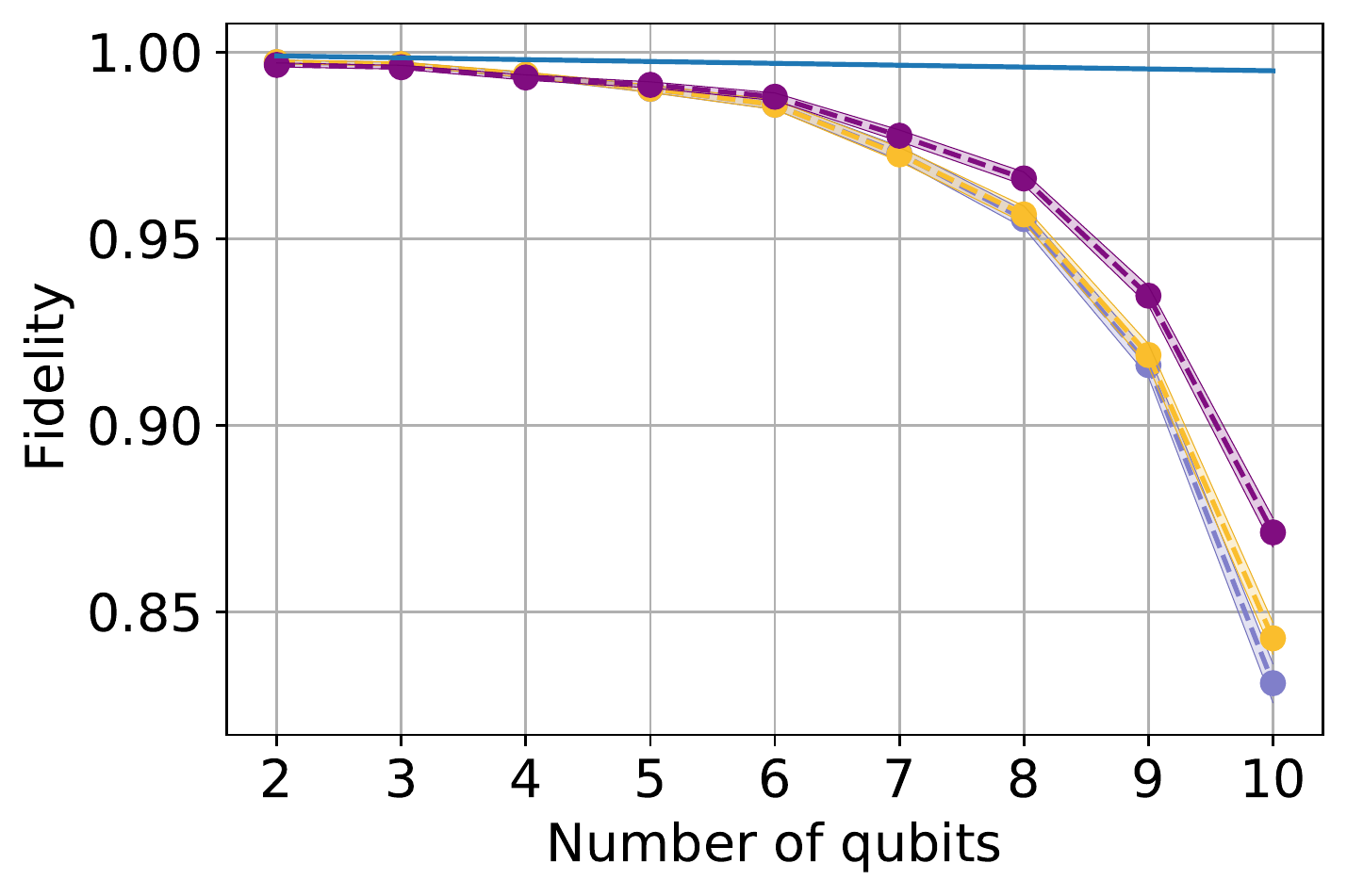}  
  \caption{$|\Phi_1^{n}\rangle$, local bases}
  \label{fig:1.1}
\end{subfigure}
\begin{subfigure}{.245\textwidth}
  \centering
  \includegraphics[width=\textwidth]{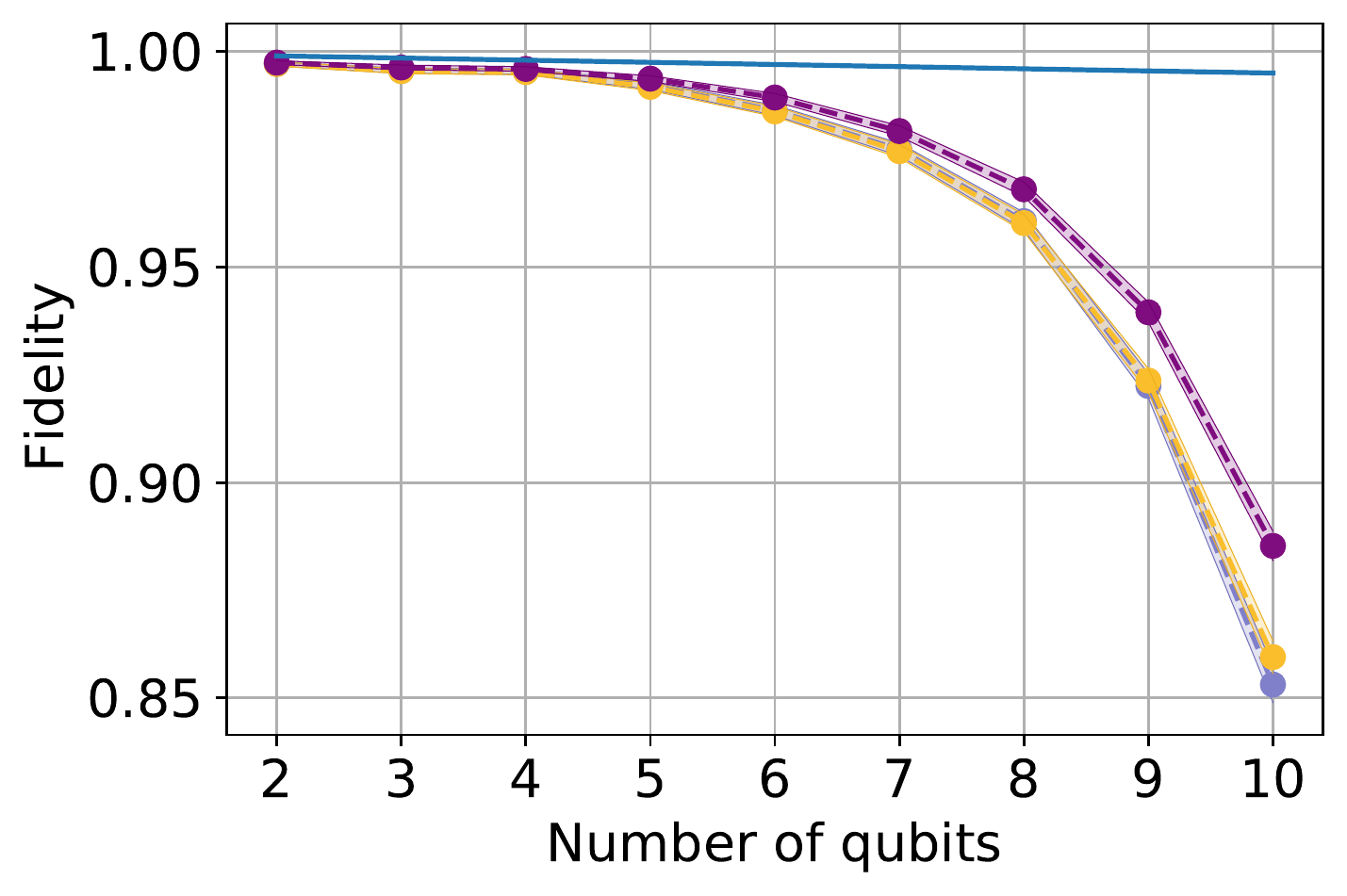}  
  \caption{$|\Phi_2^{n}\rangle$, local bases}
  \label{fig:1.2}
\end{subfigure}
\begin{subfigure}{.245\textwidth}
  \centering
  \includegraphics[width=\textwidth]{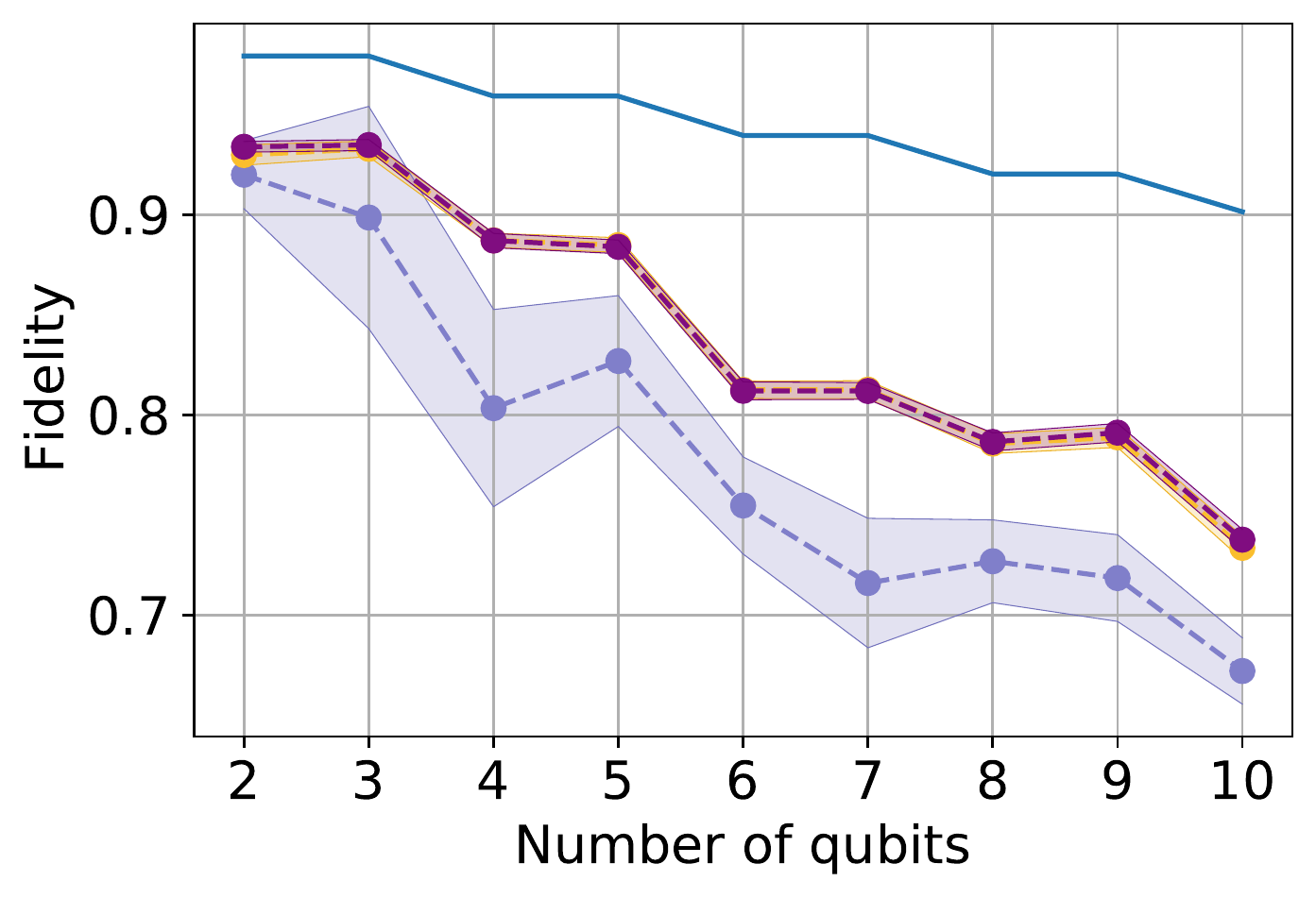}  
  \caption{$|\Phi_3^{n}\rangle$, local bases}
  \label{fig:1.3}
\end{subfigure}
\begin{subfigure}{.245\textwidth}
  \centering
  \includegraphics[width=\textwidth]{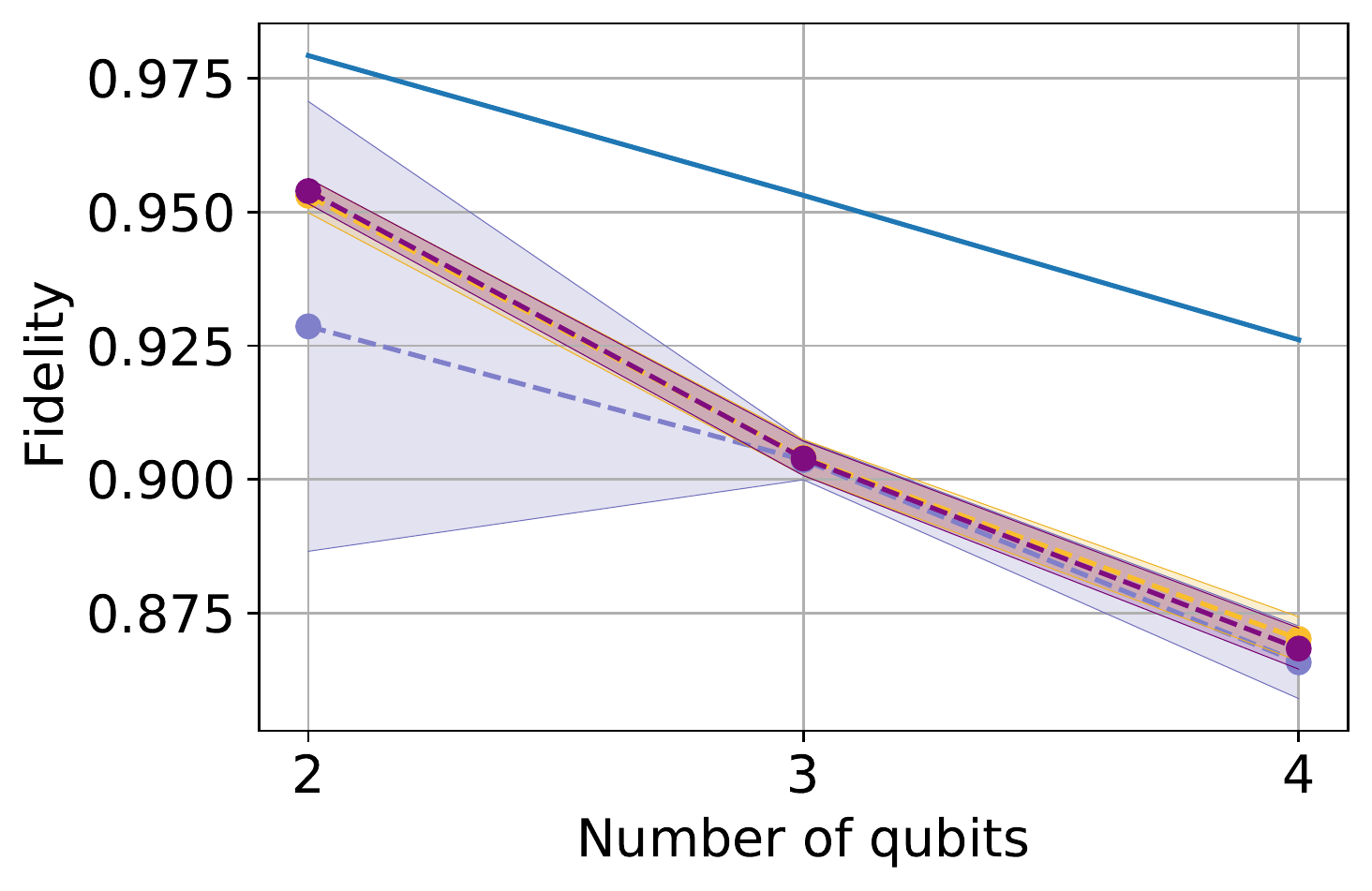}  
  \caption{$|\Phi_4^{n}\rangle$, local bases}
  \label{fig:1.4}
\end{subfigure}\\

\begin{subfigure}{.245\textwidth}
  \centering
  \includegraphics[width=\textwidth]{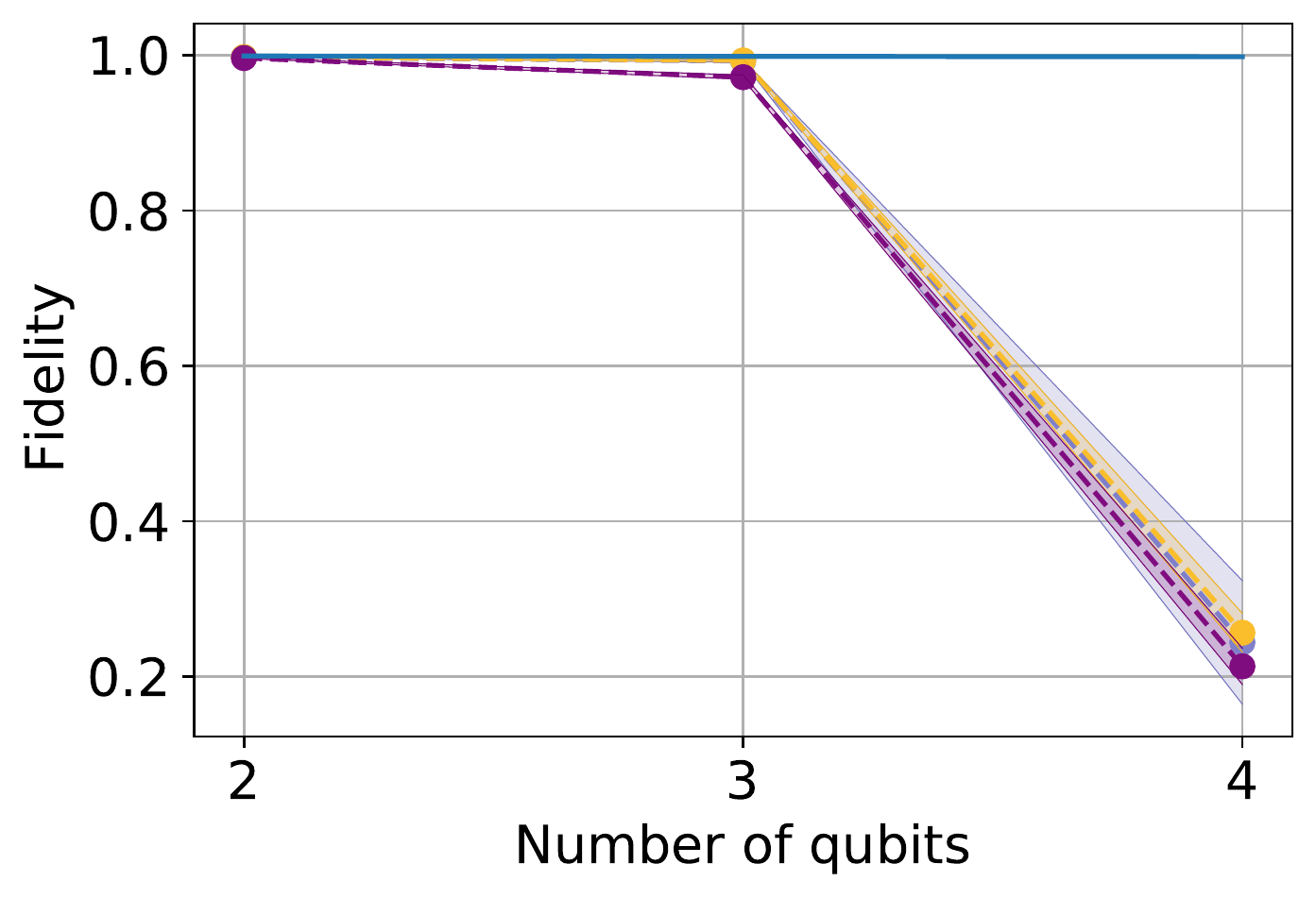}  
  \caption{$|\Phi_1^{n}\rangle$, entangled bases}
  \label{fig:2.1}
\end{subfigure}
\begin{subfigure}{.245\textwidth}
  \centering
  \includegraphics[width=\textwidth]{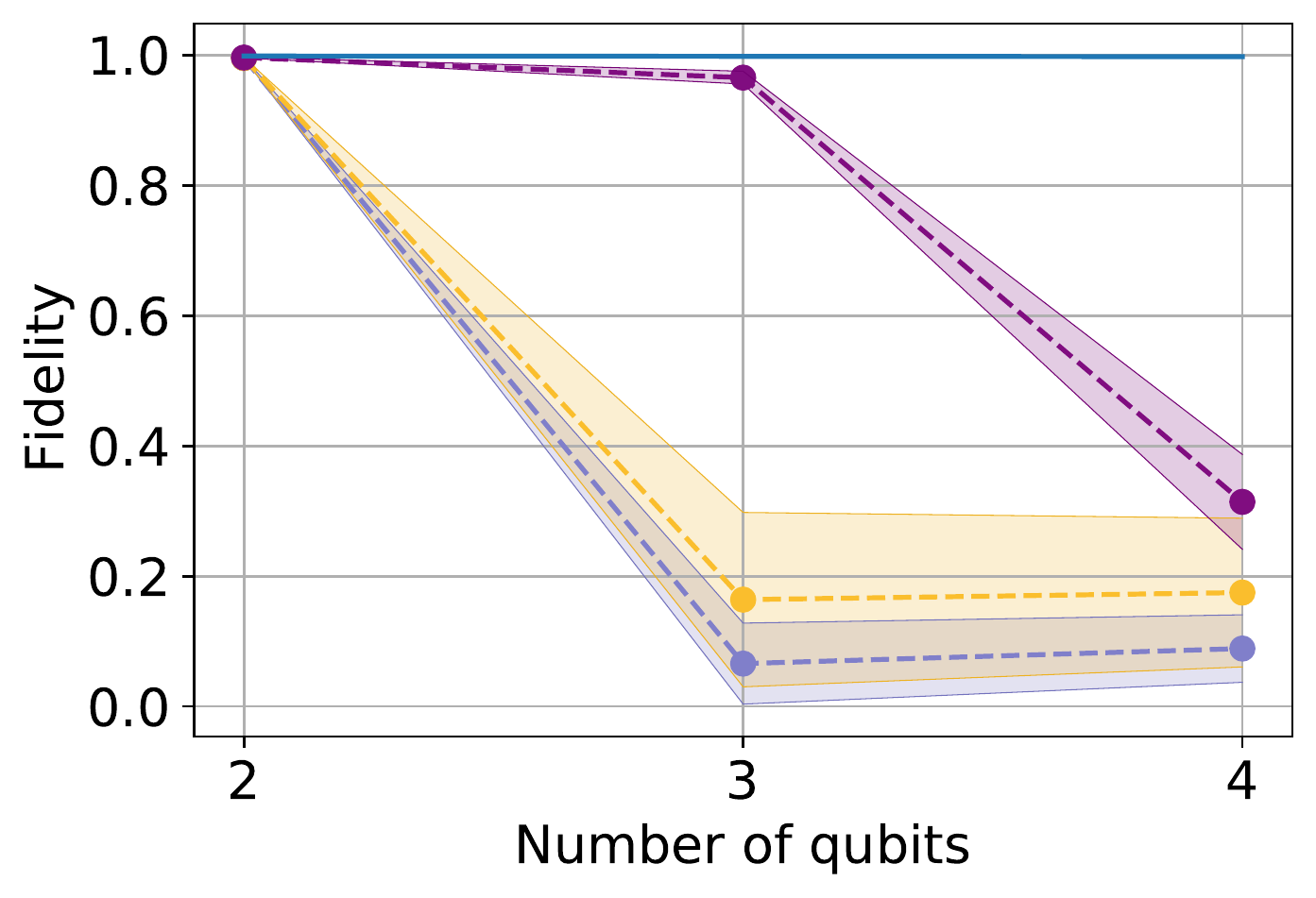}  
  \caption{$|\Phi_2^{n}\rangle$, entangled bases}
  \label{fig:2.2}
\end{subfigure}
\begin{subfigure}{.245\textwidth}
  \centering
  \includegraphics[width=\textwidth]{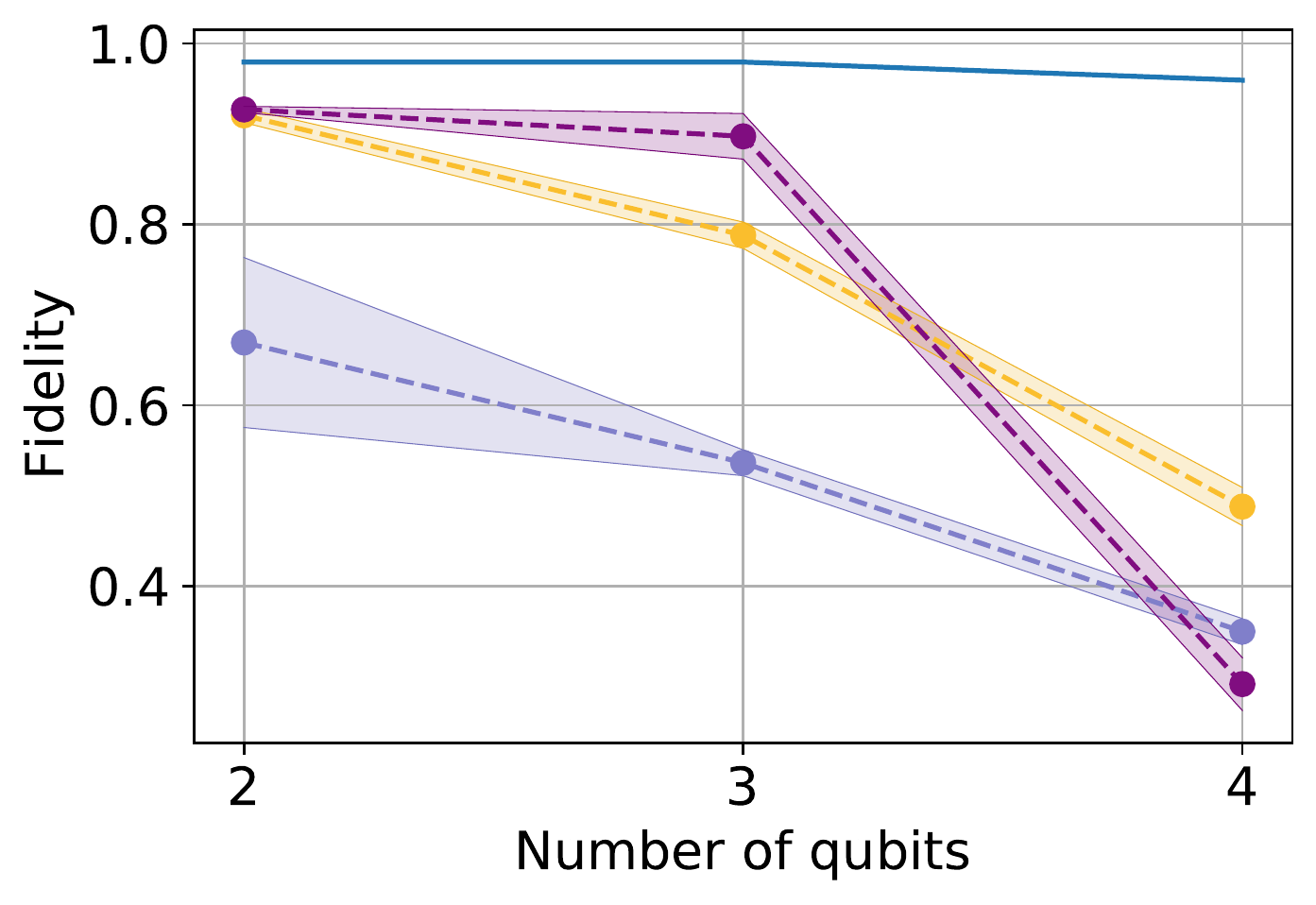}  
  \caption{$|\Phi_3^{n}\rangle$, entangled bases}
  \label{fig:2.3}
\end{subfigure}
\begin{subfigure}{.245\textwidth}
  \centering
  \includegraphics[width=\textwidth]{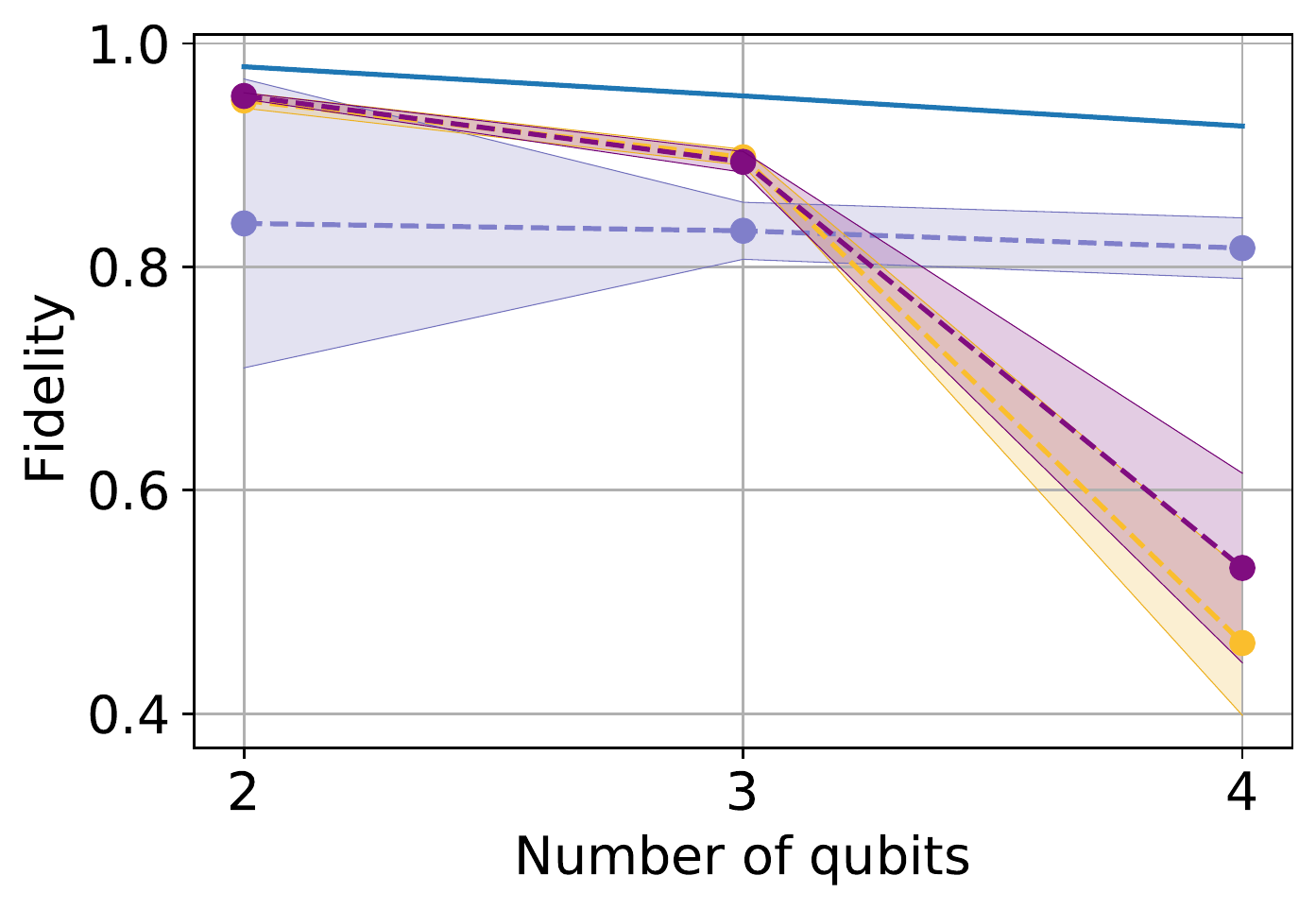}  
  \caption{$|\Phi_4^{n}\rangle$, entangled bases}
  \label{fig:2.4}
\end{subfigure}

\caption{Fidelity achieved by projective measurements on $mn+1$ local bases (upper row, $m=2$ (blue), $m=3$ (yellow), and $m=4$ (purple)) and $m$ entangled bases plus the computational base (lower row, $m=2$ (blue), $m=3$ (yellow), and $m=4$ (purple)), for states $|\Phi_1^{n}\rangle$,  $|\Phi_2^{n}\rangle$, $|\Phi_3^{n}\rangle$ and $|\Phi_4^{n}\rangle$, columns from left to right, respectively, as a function of the number $n$ of qubits.}
    \label{fig:Experiment}
\end{figure*}

\section{Experimental Realization}

We carried out an experimental demonstration of our estimation method using the quantum processor \emph{ibmq\_manhattan} developed by IBM. This NISQ device has online access and can be programmed with Qiskit, an open-source development framework for working with quantum computers in Python programming language. 

To test our estimation method we prepare the following $n$-qubit states 
\begin{align}
    |\Phi_1^{n}\rangle = \frac{1}{\sqrt{2^n}} \left( |0\rangle - e^{i\pi/4}|1\rangle \right)^{\otimes n},
\end{align}
\begin{align}
    |\Phi_2^{n}\rangle = \frac{1}{\sqrt{2^n}} \left( |0\rangle + e^{i\pi/4}|1\rangle \right)^{\otimes n},
\end{align}
and
\begin{align}
    |\Phi_3^{n}\rangle = \Bigg\{
    \begin{array}{cc}
         \frac{1}{2^{n/4}} \left( |00\rangle + |11\rangle \right)^{\otimes n/2}, & n \text{ even.}  \\
          \frac{1}{2^{(n-1)/4} } \left( |00\rangle + |11\rangle \right)^{\otimes (n-1)/2}\otimes|0 \rangle, & n \text{ odd.} 
    \end{array}
\end{align}
States $|\Phi_1^{n}\rangle$ and $|\Phi_2^{n}\rangle$ are strictly separable and thus they can be prepared efficiently with the use of a single-qubit gate acting on each qubit. The state $|\Phi_3^{n}\rangle$ has maximal entanglement between pairs of qubits. For $n$ even (odd), this state can be prepared with $n/2$ ($(n-1)/2$) CNOT gates and $n/2$ ($(n-1)/2$) single-qubit gates. Thus, states $|\Phi_1^{n}\rangle$, $|\Phi_2^{n}\rangle$, and $|\Phi_3^{n}\rangle$ can be generated by circuits with short depth. However, the state $|\Phi_3^{n}\rangle$ is generated with a lower preparation fidelity than states $|\Phi_1^{n}\rangle$ and $|\Phi_2^{n}\rangle$ due to the fact that control-not gates are implemented with an error much larger than the one achieved in the implementation of single-qubit gates.

We also perform the estimation of $n$-partite GHZ states,
\begin{align}
    |\Phi_4^{n}\rangle = \frac{1}{\sqrt{2}}\left( |0\rangle^{\otimes n}+|1\rangle^{\otimes n} \right).
\end{align}
These states are implemented by a long-depth circuit which contains $n-1$ CNOT gates. Consequently, the fidelity achieved in its generation can be very low. Thus, we focus our study in small number of qubits $n=2,3,4$ for both local and entangled bases.
Each basis was measured employing $10^{13}$ repetitions, that is, the size of the ensemble of equally prepared copies of the unknown quantum state. This is currently the maximal sample size that can be employed in IBM's quantum processors. 

Figure~\ref{fig:Experiment} summarizes the results obtained by implementing our estimation method on IBM's quantum processor, where the fidelity between states $|\Phi_1^{n}\rangle$,  $|\Phi_2^{n}\rangle$, $|\Phi_3^{n}\rangle$, $|\Phi_4^{n}\rangle$ and its corresponding estimates is displayed as a function of the number $n$ of qubits for the cases of $mn+1$ local bases (with $m=2,3$ and 4) and $m$ entangled bases plus the computational base. Shaded areas correspond to the error in the values of the fidelity obtained by bootstrapping method. The solid blue line is the maximal achievable value of the fidelity of the state preparation stage $\mathcal{\mathcal{E}}$ considering a white noise model for the device based on the average error per gate $r$ provided by IBM,
\begin{align}
    \mathcal{E}_{noise}(|0\rangle^{\otimes n}\rangle) = \left( 1-\frac{2^nr}{2^n-1}\right)\mathcal{E}(|0\rangle^{\otimes n}) + \frac{2^nr}{2^n-1}I.
\end{align}
The error model was applied on each gate necessary to prepare the states. We use as average error per gate $r_1=5\times10^{-4}$ for local gates and $r_2 =2\times10^{-2}$ for CNOT gates.

According to Figs.~\ref{fig:1.1} and \ref{fig:1.2}, the estimation of local states $|\Phi_1^{n}\rangle$ and  $|\Phi_2^{n}\rangle$ through local bases leads values that are comparable with the theoretical predictions shown in Figs.~\ref{fig:sim_2_to_10_ent} and \ref{fig:sim_2_to_10_sep}, which only consider noise due to finite statistics. For the particular case of $n=10$ qubits, the theoretical predictions for the estimation fidelity using local bases are within the intervals $[0.88, 0.93]$ for randomly generated states, and $[0.95, 0.96]$ for randomly generated separable states, while the experiment leads to an estimation fidelity in the interval $[0.82, 0.89]$.

The estimation of entangled states $|\Phi_3^{n}\rangle$ and $|\Phi_4^{n}\rangle$ by local bases, exhibited in Figs.~\ref{fig:1.3} and \ref{fig:1.4}, respectively, also leads to good fidelities, albeit lower than in the case of states $|\Phi_1^{n}\rangle$ and  $|\Phi_2^{n}\rangle$. This is to be expected since states $|\Phi_3^{n}\rangle$ and $|\Phi_4^{n}\rangle$ exhibit entanglement and thus are generated applying several CNOT gates, which increase the preparation error. Nevertheless, for the state $|\Phi_3^{n}\rangle$ with $n=10$ our method provides an estimation fidelity close to $0.67$, where the maximum achievable value according to the noise model is $0.9$ (blue solid line in Fig.~\ref{fig:1.3}). A comparison between the estimated fidelity for the bi-local state $|\Phi_3^{n}\rangle$ and GZH state $|\Phi_4^{n}\rangle$ for $n=2,3$ and $4$ qubits indicates that these entangled states are estimated with similar fidelities. This indicates that the method delivers similar quality estimation for states with different type of entanglement. The experimental realization of our estimation method shows that the use of local bases allows us to estimate pure states of large numbers of qubits. The fidelity of the estimation is meanly constrained by ensemble size and number of local bases. An increase of any of these quantities leads to an improvement in the quality of the estimation. In particular, states of larger number of qubits can be reliably estimated by increasing the number of local bases. As is shown in Figs.~\ref{fig:1.1}, \ref{fig:1.2}, and \ref{fig:1.3} the use of $4n+1$ local bases leads to higher values of the estimation fidelity than the cases of $2n+1$ and $3n+1$ local bases.

The use of entangled bases shows a different picture, where the physical realization of the method and the particular class of states to be estimated lead to a reduction of the fidelity when compared to the case of local bases. Figs.~\ref{fig:2.1} and \ref{fig:2.2} exhibit the estimation achieved for states $|\Phi_1^n\rangle$ and $|\Phi_2^n\rangle$, respectively. A comparison for $n=4$ with Figs.~\ref{fig:1.1} and \ref{fig:1.2} shows an acute decrease of the fidelity from approximately 0.995 to 0.2. A similar decrease can also be observed in the case of states $|\Phi_3^n\rangle$ and $|\Phi_4^n\rangle$. The low achieved fidelity finds its origin in the circuits employed to implement the measurements on the entangled bases. For $n=2,3$, and $4$, 2 entangled bases require the use of 2, 7 and 27 control-note gates, respectively. This number becomes 3027 for $n=10$. Since the control-not gate has a high error rate in NISQ-devices, the implemented bases differ significantly from the actual bases $\mathcal{E}_a$ to be implemented. Besides, this also affects the generated states. Fig.~\ref{fig:2.2} shows that for $n=3$ the use of $2$ and $3$ entangled bases plus the computational base lead to a fidelity value of $0.1$ much lower than the case of using $4$ entangled bases plus the computational base. This originates in the state $|\Phi_2^3\rangle$ to be reconstructed, which for $2$ and $3$ entangled bases exhibits ill-conditioned equation systems. This is not present in the case of $4$ entangled bases, where the fidelity is in the order of $0.95$. Let us note that this effect does not appear in Fig. ~\ref{fig:1.2}, where the use of $mn+1$ local bases allow us to obtain well-conditioned equation systems. Ultimately, the combinations of these factors leads to a low estimation fidelity. 

\section{Conclusions}

Estimating states of $d$-dimensional quantum systems requires a minimal number of measurement outcomes that scales quadratically with the dimension. In the case of composite systems, such as quantum computers, the scaling becomes exponential in the number of subsystems, which makes the estimation by generic methods unfeasible except for few-component systems. NISQ computers are characterized by low-accuracy entangling gates, which are required for implementing measurements of arbitrary observables, and by a fixed number of repetitions, which constrains the size of statistical samples and decreases the estimation accuracy as the number of qubits increases. Therefore, estimating $n$-qubit states of NISQ computers is difficult task. 

We have proposed a method to estimate pure states of $n$-qubit systems that is well suited for NISQ computers. The method is based on the reconstruction of the so-called reduced states $|\tilde\Psi_{\beta}^j \rangle$, which for an unknown state $|\Psi \rangle$ are $n$ sets of $2^{n-j}$ non-normalized states, one set for each $j=1, 2, ..., n$. If we know the reduced states $|\tilde\Psi_{\beta}^j \rangle$ for a fixed $j$, then it is easy to reconstruct the reduced states with $|\tilde\Psi_{\beta}^{j+1} \rangle$ from the results of well defined set of measurements. The method begins by measuring a set of projectors that allow us to reconstructs all the reduced states for $j=1$, and iterating, the rest of reduced states until reaching $|\tilde\Psi_{\beta}^n \rangle$, the estimate of the unknown pure state. 

The set of measurements employed by the proposed method corresponds to projective measurements onto a set of bases. We have first shown that $mn+1$ bases, with $m$ an integer number equal or greater than 2, allow us to estimate most pure states up to a null-measure set. Thereby, a total of $(mn+1)2^n$ measurement outcomes is required. This number compares favorably with other estimation methods. Mutually unbiased bases , SIC-POVM, and compressed sensing require $(2^n+1)2^n$, $2^{2n}$, and in the order of $2^{2n}n^2$ measurement outcomes, respectively, to reconstruct pure states. The $mn+1$ bases are local, that is, they can be cast as the tensor product of $n$ single-qubit bases. Projective measurements onto the states of these bases are carried out on a NISQ computer by applying local gates onto each qubit followed by a projection onto the computational base. Thus, no entangling gates are necessary. In principle, $2n+1$ bases are enough. We have shown, by means of numerical simulations, that the use of a larger number of local bases increases the estimation accuracy for larger numbers of qubits. However, if high-accuracy entangling gates are available, then the set of measurements employed by the proposed method can be reduced to projective measurements on $m$ entangled bases plus the computational base, with $m$ an integer number equal or greater than 3. In this case, an increase in the number of entangled bases also allows us to increase the estimation accuracy.

We tested the proposed estimation method in the IBM's quantum processing units. The estimation of local pure states via local bases up to 10 qubits provides fidelities that agree with the numerical simulations and are above 90\%. The estimation of entangled states via local bases was also tested. In this case, the estimation of tensor product of two-qubit Bell states up to $n=10$ led to fidelities above 70\%. In this case, however, the preparation of the state is affected by large errors due to the use of entangling gates. Finally, we tested the estimation of a GHZ state for $n=2,3$ and 4 qubits obtaining fidelities above 86\%. Entangled bases were also tested. Nevertheless, due to the massive use of entangling gates in the preparation and measurement stages very low fidelities were achieved, as expected. The estimation via $k$ entangled bases plus the computational base, where $k$ is high-enough but does not scale with the number $n$ of qubits, can lead to a large fidelity in future fault-tolerating quantum architectures \cite{Knill}, where a large numbers of qubits and high-accuracy entangling gates are required.
Also, it might be possible to increase the estimation fidelity by using alternative implementations of the multi-control NOT gates that employ ancillary qubits. This allows a reduction in the depth of the circuit required to implement measurements on the proposed entangled bases \cite{MultiControlGates}.

The main characteristic of our protocol is its scalability, which is a consequence of the {\it a priori} information about the states to be estimated. This is an approach common among many estimation methods such as, for instance, compressed sensing and matrix product states. The purity assumption is reasonable in systems that are able to prepare high purity states or in systems whose purity can be certified, for instance, through randomized benchmarking \cite{RandomizedBenchmark}. However, this is not necessarily true on NISQ devices. Decoherence and gate errors can reduce the quality of preparing a pure state, so it really becomes a mixed state. In this case our protocol could not be applied. Nevertheless, it has been shown that experimental results can be improved by reducing the error of the raw data employing error-mitigation techniques \cite{CI5BB,noise1,error_mitigation_1,error_mitigation_2}. Combining scalable error-mitigation methods with our estimation method could extend its applicability to high-noise systems.

\begin{acknowledgments}
This work was supported by ANID -- Millennium Science Initiative Program -- ICN17$_-$012. AD was supported by FONDECYT Grant 1180558. LP was supported by ANID-PFCHA/DOCTORADO-BECAS-CHILE/2019-72200275. LZ was supported by ANID-PFCHA/DOCTORADO-NACIONAL/2018-21181021. We thank the IBM Quantum Team for making multiple devices available to the CSIC-IBM Quantum Hub via the IBM Quantum Experience.
\end{acknowledgments}


\end{document}